**Large magnetic moment in flexoelectronic silicon at room temperature**

Paul C. Lou[1‡], Anand Katailiha[1‡], Ravindra G. Bhardwaj[1‡], Ward P. Beyermann[2], Dominik M. Juraschek[3], and Sandeep Kumar[1,4,*]

[1]Department of Mechanical engineering, University of California, Riverside, CA 92521, USA

[2] Department of Physics and Astronomy, University of California, Riverside, CA 92521, USA

[3] Harvard John A. Paulson School of Engineering and Applied Sciences, Harvard University, Cambridge, MA 02145, USA

[4] Materials Science and Engineering Program, University of California, Riverside, CA 92521, USA

**Abstract**

Time-dependent rotational electric polarizations have been proposed to generate temporally varying magnetic moments, for example, through a combination of ferroelectric polarization and optical phonons. This phenomenon has been called dynamical multiferroicity, but explicit experimental demonstrations have been elusive to date. Here, we report the detection of a temporal magnetic moment as high as 1.2 $\mu_B$/atom in charge-doped thin film of silicon under flexural strain. We demonstrate that the magnetic moment is generated by a combination of electric polarization arising from a flexoelectronic charge separation along the strain gradient and the deformation potential of phonons. The effect can be controlled by adjusting the external strain gradient, doping concentration and dopant, and can be regarded as a dynamical multiferroic effect involving flexoelectronics polarization instead of ferroelectricity. The discovery of a large magnetic moment in silicon may enable the use of non-magnetic and non-ferroelectric semiconductors in various multiferroic and spintronic applications.

Magnetoelectric multiferroic materials[1, 2] allow control of the spin and the charge in electric and magnetic fields[3], respectively. There have been very few magnetoelectric ferroelectric materials that led to research in strain engineering, interface control, nanostructuring and heterostructures to achieve large magnetoelectric coupling.[1, 2, 4] Similar research effort has been undertaken using the inhomogeneous magnetoelectric effects[5] that arise due to the spatial magnetic inhomogeneity in a magnetic crystal.[6, 7] However, ferroelectric materials were essential in both cases. Recently, an alternate mechanism, called dynamic multiferroicity, was proposed to achieve magnetic order in ferroelectric materials.[8, 9, 10] The proposed dynamical multiferroic response results from a combination of the ferroelectric polarization with the dynamical (time–dependent) polarization from optical phonons and moving ferroelectric domain walls.[10, 11] The resulting magnetic moment resulting from dynamical polarization can be described by the following relationship-

$$M_t \sim P \times \partial_t P \qquad (1)$$

where M and P are respectively magnetic moment and polarization. The electric polarization can hereby be given through different origins, such as ferroelectricity and optical phonons. Recently, Bliokh et al[12] proposed electric-current-induced non-reciprocal propagation of surface plasmon-polaritons, which could be considered a case of electronic dynamical multiferroicity. Similarly, the flexoelectric polarization coupled with phonons (time dependent polarization) can give rise to a dynamical magnetic moment and a magnetoelectric response in any material system, which will allow control of spin, charge and phonon degree of freedom in technologically important materials (semiconductors).

While the flexoelectric polarization is usually orders of magnitude smaller than ferroelectric polarization, Yang et al[13] recently, demonstrated a large flexo-photovoltaic effect in centrosymmetric materials including silicon. Similarly, a very large flexoelectronic effect[14] has been reported in centrosymmetric materials including Si. The observation of a large flexoelectronic effect in silicon was attributed to the electronic-charge separation due to strain gradient, as shown in Figure 1 (a). This flexoelectronic response arises due to the agglomeration of electronic charge in the region with tensile strain and loss of electronic charge in the region with compressive strain.[15, 16, 17] In addition, the charge carrier separation caused by strain gradient changes the band gap and also contributes to the flexoelectronic response. The magnitude of this flexoelectronic polarization in silicon can be compared to the magnitude of the polarization in ferroelectric materials,[15] which led us to choose Si for our study. Recently, a large spin-Hall effect(SHE)[18] has also been reported in silicon that is attributed to strain gradient mediated spin-phonon coupling. These reports indicated that inhomogeneous strain can give rise to spin, charge and phonon entanglement in centrosymmetric materials. While coupling between the charge carrier and phonons can arise due to superposition of flexoelectronics charge separation and phononic deformation potential, the origin of the spin component needed for SHE is unclear. This motivated us to explore the inhomogeneous strain mediated spin, charge and phonon coupling behavior in centrosymmetric silicon.

Here, we took angle-dependent magneto-transport measurements on p-doped and n-doped silicon. We found that the flexoelectronic charge separation coupled with the transverse optical (TO) and transverse acoustic (TA) phonons gives rise to a magnetic moment of 1.2 $\mu_B$/atom, aligned primarily along the $<111>$ tetrahedral bonding directions in diamond cubic silicon lattice. Phenomenologically, the symmetry of the dynamical multiferroic effect in silicon is identical to the theoretically predicted symmetry of dynamical multiferroic effect, as shown in Figure 1 (b). The dynamical multiferroic effect gives rise to entanglement of spin, charge and phonon in silicon (as shown in Figure 1 (b)), which is the underlying cause of phonon skew scattering mediated SHE and anomalous Hall effect (AHE) in silicon.

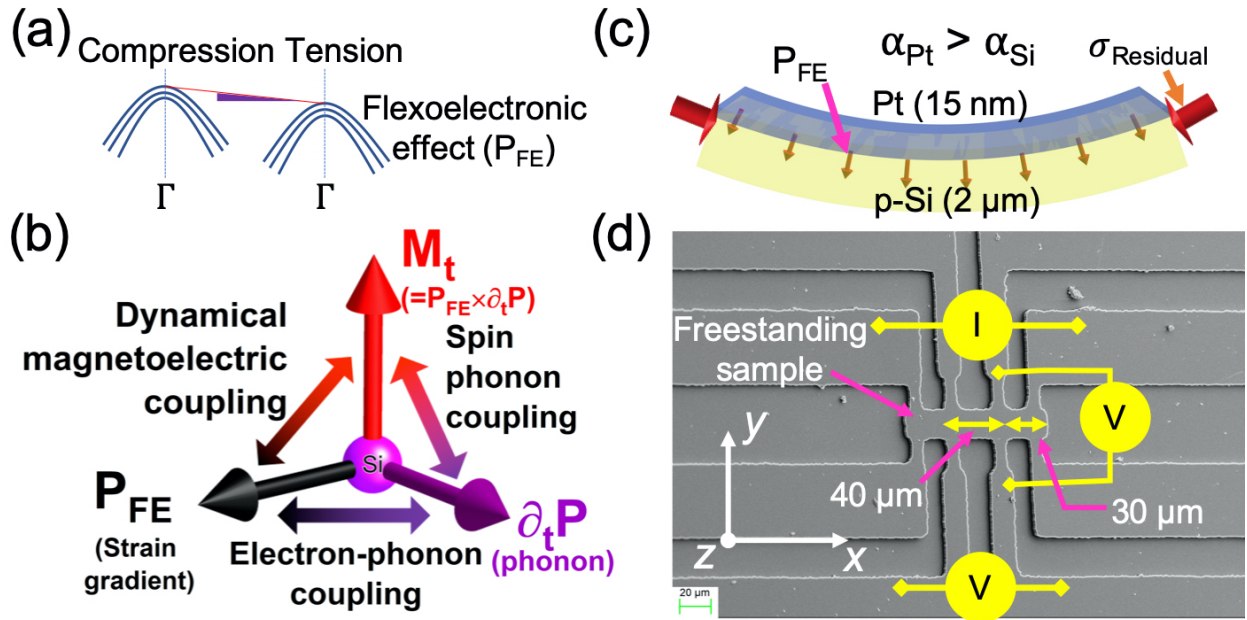

Figure 1. (a) a schematic showing the change in valence band structure due to strain gradient, which gives rise to a charge separation and a large flexoelectronic field, (b) a schematic showing the dynamical multiferroic effect that is expected to give rise to spin, charge and phonon entanglements, (c) a representative scanning electron micrograph showing the structure of sthe ample and measurement scheme for longitudinal and transverse responses, (d) a schematic showing the buckled thin film structure caused by residual stresses (thermal mismatch) and giving rise to flexoelectronic (FE) polarization.

In the first experiment, we fabricated (materials and methods) an experimental setup with a freestanding p-silicon sample, as shown in Figure 1 (c) and Supplementary Figure S1. To induce a large strain gradient, we deposited 1.8 nm of MgO and 15 nm Pt on top of the p-silicon layer. The melting point of Pt is very high compared to silicon and thin film deposition using e-beam evaporation will lead to large residual thermal mismatch stresses. A similar technique has been used to demonstrate a second harmonic generation in silicon waveguides[19] using silicon nitride over layer. The buckling of the composite sample and, as a consequence, strain gradient as well as flexoelectronic polarization arises in p-silicon as shown in Figure 1 (d).[18] The resistivity of the Pt and p-silicon layers (p-silicon wafer resistivity of $1.0×10^{-5}$ - $5×10^{-5}$ $\Omega$-m) were $2.02×10^{-7}$ $\Omega$-m and $1.2×10^{-5}$ $\Omega$-m, respectively. In this bilayer (effectively) configuration, approximately 69.17% of the current will pass through the p-silicon layer. The Pt layer is metallic and the highly doped p-silicon layer also shows metal-like behavior.[20] However, the composite sample had a complex resistance behavior as a function of temperature, as shown in Figure 2 (a). To understand the temperature-dependent resistance behavior, we measured the angle-dependent (zy-plane) transverse resistance as a function of constant magnetic field and temperature.

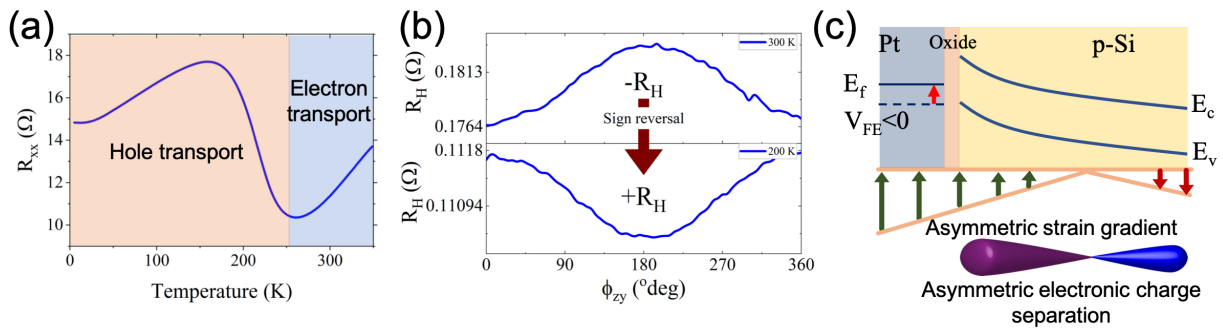

Figure 2. (a) the longitudinal resistance measurement as a function of temperature, (b) angle-dependent transverse resistance as a function of magnetic field (1 T) and temperature (300 K and 200 K) showing the sign reversal of Hall resistance, and (c) the expected band structure due to asymmetric strain gradient and asymmetric charge

separation in the sample that give rise to the flexoelectronic effect and the sign reversal of Hall resistance.

The angle-dependent transverse resistance is negative at 300 K, corresponding to the electrons as the charge carrier, as shown in Figure 2 (b). However, at 200 K, the angle-dependent transverse resistance exhibits a sign reversal, as shown in Figure 2 (b), which indicates a hole mediated charge transport. Since Pt and p-silicon thin have electrons and holes as charge carriers, respectively; a sign change could occur because of their relative contributions to Hall resistance. However, the averaged charge carrier concentration is estimated to be $9.09 \times 10^{20}$ cm$^{-3}$ and $4.88 \times 10^{21}$ cm$^{-3}$ at 300 K and 200 K, respectively, which is two orders of magnitude larger than the expected carrier concentration in p-silicon (~$4 \times 10^{19}$ cm$^{-3}$). Hence, the sign reversal of Hall resistance can only come from changes in the Fermi energy and band structure in the Pt layer. In our composite sample, the reduction in temperature increases the thermal mismatch stresses and strain gradient. Consequently, a large flexoelectronic polarization at the interface acts as a gate bias and changes the Fermi level in the Pt layer causing a transition to a hole-mediated charge transport, as shown in Figure 2 (b),[14] similar to the ionic gating of Pt reported by Liang et al.[21] The large flexoelectronic effect results from the gradient in the band structure due to the strain gradient and the consequent electronic charge separation, as shown in Figure 2 (c).[15] Based on the resistance measurement presented in Figure 2 (a), the charge transport in the Pt layer is expected to transition from electrons to holes at a temperature lower than ~250 K, which is the underlying cause of the increase in resistance at that temperature. In the highly doped silicon sample, the screening caused by free charge carriers should extinguish the flexoelectronic effect, which did not happen. We attribute the large flexoelectronic effect in highly doped samples to the asymmetric strain gradient, as shown in Supplementary Figure S2. The free charge carriers do not screen the electronic charge density in the case of the asymmetric strain gradient as shown in Figure 2 (c).

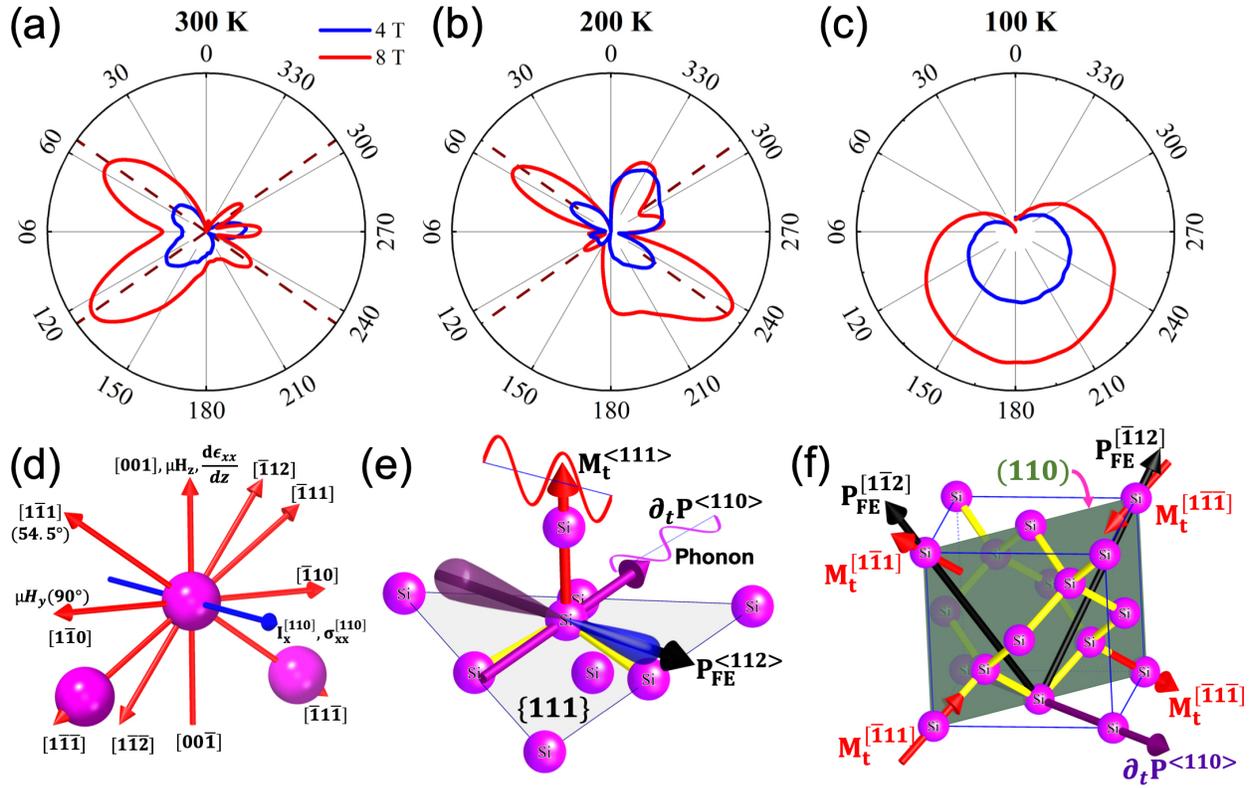

Figure 3. The polar plot of angle dependent (in the zy-plane) magnetoresistance (magnitude) in the Pt/p-silicon sample as a function of the constant magnetic field of 4 T and 8 T at (a) 300 K and (b) 200 K and (c) 100 K; the dotted lines represent $<111>$ directions. (d) The primary crystallographic directions (red color) in the zy- or (110) plane in the silicon sample observed in the measurement and direction of the current and stress (blue color), (e) a schematic showing the origin of dynamical multiferroic behavior and the temporal variation in magnetic moment due to coupling between flexoelectronic polarization and dynamical polarization caused by phonons, and (f) a schematic of the silicon lattice showing the flexoelectronic polarization and the observed magnetic moments in the (110) cross-sectional plane caused by dynamical multiferroic effect.

Next, we acquired and analyzed the angle-dependent longitudinal resistance as a function of the magnetic field in the zy-plane (the plane containing the strain gradient) where the field was always perpendicular to the direction of the current and the strain. We found a relatively large MR (12.7 %) at 300 K when the magnetic field was at 8 T, as shown in Supplementary Figure S3. This MR behavior is not observed in either p-silicon only thin films[22] or Pt thin films. However, the MR reduces significantly and almost disappears as the temperature is reduced to 5 K, as shown in Supplementary Figure S3. A closer analysis of the magnitude of the angle-dependent MR (ADMR) measurement shows a crystallographic direction-dependent anisotropy for an applied magnetic field of 4 T and 8 T at 300 K and 200 K, as shown in Figure 3 (a, b). However, the anisotropic behavior disappears at 100 K, as shown in Figure 3 (c). The raw ADMR responses are shown in Supplementary Figure S4. It is noted that the direction-dependent behavior is not clearly visible at 300 K and at an applied magnetic field of 1 T, as shown in Supplementary Figure S5. The cross-section of the silicon sample is the (110) plane (zy-plane) and the vertical direction (z-axis) is identified to be [001], as shown in Figure 3 (d). As a consequence, the lowest resistance was observed when the magnetic field was pointed along either [1$\bar{1}$1] or [1$\bar{1}\bar{1}$] directions (at 54.7° from ±z- axis) at both 300 K and 200 K. The modulations in the ADMR were also observed for angles corresponding to the [1$\bar{1}$0] and [1$\bar{1}\bar{2}$] directions, respectively.

The ADMR response suggests that a local magnetic moment in the silicon is preferentially aligned along the primary symmetry directions of the diamond cubic lattice, which we propose arises due to the dynamical multiferroic effect. The electronic charge separation gives rise to anisotropic flexoelectronic behavior along the principal symmetry directions of the lattice[15] due to the linear atomic density variations, as shown in Supplementary Figure S6, which is 41.4% larger along $<110>$ directions as compared to the $<100>$ directions. However, in this study, there are no $<110>$ directions in the (110) cross-sectional plane of the sample. The four $<110>$ directions, having out of plane components, give rise to the local flexoelectronic charge separation anisotropy along $[1\bar{1}2]$ and $[\bar{1}12]$, as shown in Supplementary Figure S6. The direction of the current is along [110] direction and, as a consequence, temporal polarization is also expected to be along [110] direction. The $P_{FE}^{<112>}$ and $\partial P_t^{<110>}$ leads to the temporal magnetic moment of $M_t^{<111>}$ as shown in Figure 3 (e). In our samples, the $P = \left(P_{FE}^{[1\bar{1}2]}, \partial P_t^{[110]}, 0\right)$ gives rise to $M_t^{[\bar{1}11]}$ and $M_t^{[1\bar{1}\bar{1}]}$ as shown in Figure 3 (f). Similarly, the $P = \left(P_{FE}^{[\bar{1}12]}, \partial P_t^{[110]}, 0\right)$ gives rise to $M_t^{[1\bar{1}1]}$ and $M_t^{[\bar{1}1\bar{1}]}$ as shown in Figure 3 (f). Hence, the symmetry of the response is consistent with the dynamical multiferroicity. The optical phonons, expected to be responsible for dynamical multiferroicity, in silicon carry insignificant polarization, and the resulting temporal magnetic moment should also be small. However, superposition of the flexoelectronic charge separation and the phononic deformation potential[23] can potentially lead to the formation of charged phonons[24], which consequently can give rise to a large temporal magnetic moment. More importantly, the crystallographic direction-dependent behavior disappears at 100 K and below, as shown in Figure 3 (c) and Supplementary Figure S3 as expected for a phonon mediated behavior. The symmetry of the response and the temperature dependent behavior strongly suggests that the dynamical multiferroic effect gives rise to the temporal magnetic moment, which is possibly the underlying reason for the observed ADMR response.

The flexoelectronic charge separation, the phononic deformation potential and the temporal magnetic moment combined to give rise to the spin dependent electron-phonon interactions and the spin-phonon coupling[25] in our samples, as shown in Figure 1 (b). In the ADMR response measurement, the externally applied magnetic field polarizes the spin of the charge carriers. Consequently, spin-dependent electron-phonon interactions gives rise to modulations in the resistance as a function of the crystallographic direction. The spin dependent scattering decreases whenever the magnetic field is aligned parallel to the direction of the temporal magnetic moment ($M_t^{<111>}$). However, we needed additional experimental evidence for the spin-phonon coupling and the dynamical multiferroic effect. The spin-phonon coupling also gives rise to phonon skew scattering[26] and SHE. But, such a SHE will diminish as a function of temperature, as shown by Karnad et al[27] unlike the SHE due to spin-orbit coupling.

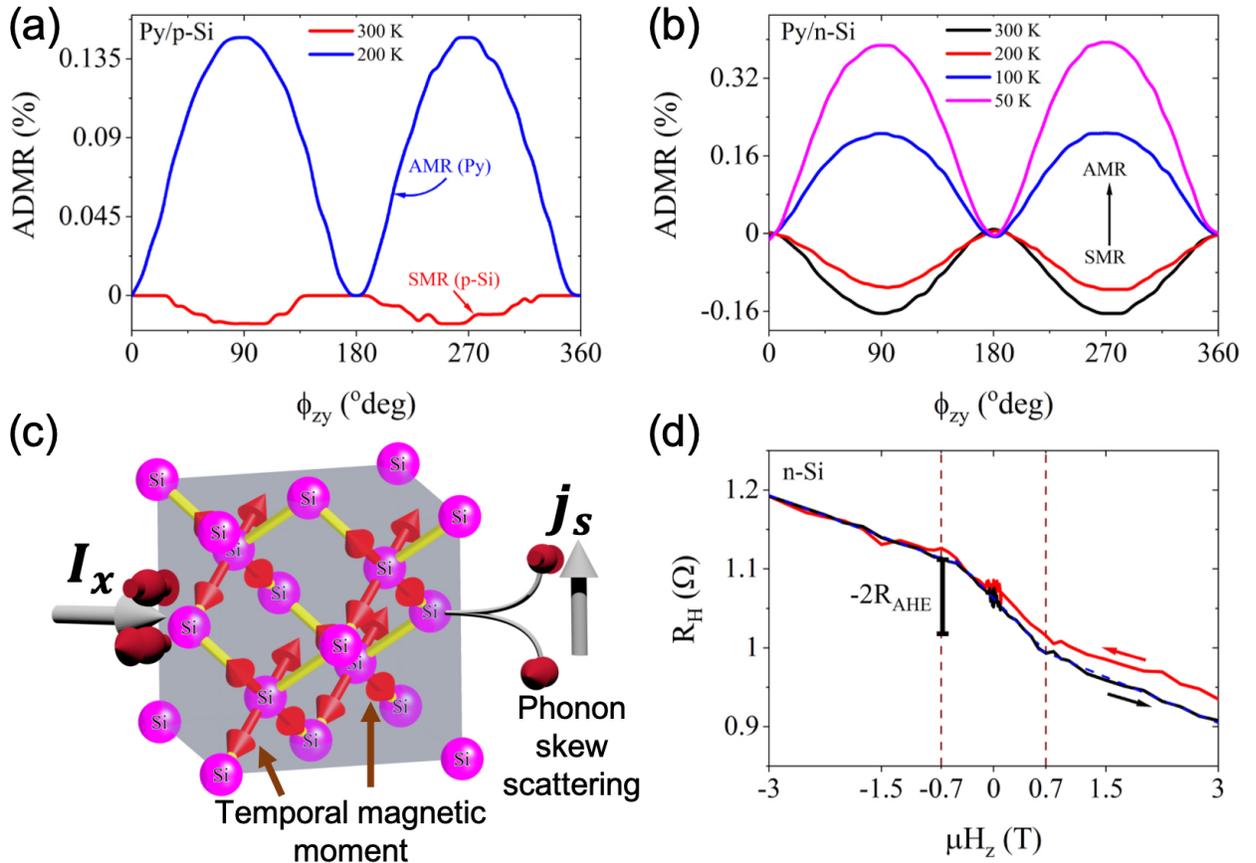

Figure 4. The ADMR response in the zy-plane at the 4 T magnetic field and as a function of the temperature showing the changes in SMR behavior in (a) the Py/p-silicon sample

and (b) the Py/n-silicon sample. (c) A schematic showing the temporal magnetic moment along $<111>$ directions due to the dynamical multiferroic effect that caused the phonon skew scattering and the resulting transverse spin current in the Si and (d) the anomalous Hall effect measurement in the n-silicon sample at 5 mA of current.

To discover the SHE behavior as a function of temperature, we chose a technique of spin-Hall magnetoresistance (SMR) measurement. In the SMR measurement, a ferromagnetic secondary layer leads to modulations in the resistance due to the relative orientation of the spin polarization from SHE and the magnetic moment in the ferromagnetic layer.[28, 29] We chose Py ($Ni_{80}Fe_{20}$- permalloy) as a ferromagnetic layer for our measurements. The SMR response as a function of temperature in p-silicon and n-silicon samples were measured at the 4 T applied magnetic field and the sample rotation was done in the zy-plane. Similar to the Pt thin film in the first sample, the residual stresses due to the Py layer also induce asymmetric strain gradient and break the inversion symmetry in Si. At 300 K, the response in the zy-plane included SMR ($-\sin^2\theta$ symmetry) from the p-silicon (2 µm) layer and anisotropic magnetoresistance (AMR) ($+\sin^2\theta$ symmetry) from Py (25 nm) layer, as shown in Figure 4 (a).[18] The total response was $1.5\times10^{-4}$ at 300 K but the SMR behavior disappeared at 200 K in the p-silicon sample, as shown in Figure 4 (a). In the Py (75 nm)/n-silicon (2 µm) sample, the magnitude of the SMR response decreases from $1.64\times10^{-3}$ at 300 K to $1.09\times10^{-3}$ at 200 K and then disappears at 100 K, as shown in Figure 4 (b). At lower temperatures, only the AMR response from the Py layer remains in both samples. The magnitude of the SMR response is an order of magnitude larger than that of Pt,[30] and it is of the same order as the SMR reported in some topological insulators.[31, 32] Since the SMR response disappears at lower temperatures, the phonon skew scattering due to spin-phonon coupling is the underlying cause of large SMR in the Py/silicon composite samples, as shown in Figure 4 (c). The phonon skew scattering behavior in our samples is caused by temporal magnetic moment rather than spin-orbit coupling.

From the ADMR behavior shown in figure 3 (a,b), we observe that the magnetic moment along $<111>$ directions can lead to a net magnetic moment due to asymmetry. If the asymmetry of the temporal magnetic moment is negligible, the SHE response will be observed whereas a large asymmetry will lead to AHE because the underlying mechanism for both SHE and AHE is same. The AHE response can allow us to estimate the magnitude of the temporal magnetic moment. In addition, the samples discussed so far had a metal layer that caused asymmetric strain gradient. We needed to induce an asymmetric strain gradient in the silicon layer without using the metal layer deposition to eliminate the contribution of the Pt and Py layers. Previous studies[20] have demonstrated that asymmetry in silicon can be induced using 2 nm of MgO layer due to large residual stresses. We made the p-silicon and n-silicon samples with 2 nm of MgO to uncover the dynamical multiferroicity. Then, we measured the Hall resistance for the p-silicon and n-silicon samples to uncover the AHE behavior. We did not observe any AHE response in the p-silicon sample, as shown in Supplementary Figure S7. But, the p-silicon sample shows a negative Hall resistance, which is attributed to the flexoelectronic mediated charge separation. In contrast, the n-silicon sample shows a negative AHE response of 34.13 m$\Omega$ for a 5 mA of AC bias, as shown in Figure 3 (d). It is noted that the AHE response is almost non-existent in the n-silicon sample at 1 mA of AC bias, as shown in Supplementary Figure S8. At a higher current, the thermal expansion stresses are larger in the freestanding sample leading to increased buckling strain gradient and larger flexoelectronic charge separation, which is supported by the smaller charge carrier concentration of ~5.9×10$^{19}$ cm$^{-3}$ at 1 mA of current as compared to ~8×10$^{19}$ cm$^{-3}$ at 5 mA. Similarly, the resistivity of the sample also reduces from ~2.44×10$^{-5}$ Ω-m (34.95 Ω) at 1 mA to ~1.5×10$^{-5}$ Ω-m (24.73 Ω) at 5 mA. The reduction in resistivity is more than the increase in the charge carrier concentration, which is attributed to the increase in mobility as well as the reduction in scattering possibly due to the larger flexoelectronic charge separation.

The saturation magnetization is estimated to be ~0.7 T, as shown in Figure 4 (d) and the magnetic moment was calculated to be 1.2 µ$_B$/atom or 9.6 µ$_B$/cell. The temporal magnetic moment is orders of magnitude larger than the one predicted in dynamical multiferroicity. We attribute this large temporal magnetic moment to electron-phonon coupling. The large electron-phonon coupling arises due to superposition of the flexoelectronic charge separation and the deformation potential due to phonons, as stated earlier. The dynamical multiferroic response in our samples is an electronic effect rather than ionic. Consequently, the temporal magnetic moment is orders of magnitude larger. Cheng et al[33] reported a large phononic magnetic moment (2.7 µ$_B$) in the Dirac semimetal Cd$_3$As$_2$, which they also attributed to the electron-phonon coupling constant. The large temporal magnetic moment is the underlying cause of large SMR response, especially in n-silicon samples. The TA phonons are most likely to give rise to spin dependent scattering. This assertion is supported by recently reported magnetic field dependent thermal conductivity behavior.[22] That study also reported optical phonon softening due to spin-phonon coupling using Raman measurements.[22] Hence, the deformation potential of both TA and TO phonons will couple to the flexoelectronic charge and temporal magnetic moment in our samples.[23] The measurements in this study are quasi-static and information regarding the time-dependent behavior of the magnetic moment cannot be extracted.

In conclusion, we presented the possible experimental evidence of a large magnetic moment of 1.2 $\mu_B$/atom or 9.6 $\mu_B$/cell in silicon that we think arises from superposition of flexoelectronic charge separation and phonon deformation potential, which can be regarded as a dynamical multiferroic effect. In addition to multiferroicity, the flexoelectronic response also gives rise to strong electron-phonon and spin-phonon coupling, which manifests as spin-dependent phonon skew scattering and gives rise to a large spin-Hall effect and an anomalous-Hall effect. This work demonstrates that spin, charge and phonon entanglement can be tailored using inhomogeneous strain even in centrosymmetric materials and materials with weak intrinsic spin-orbit coupling. Our results facilitate novel ways to use nonmagnetic and non-ferroelectric semiconductors in different spin-based applications, involving spintronics and spin-caloritronics, as well as magnetoelectric devices.

**Corresponding author**

*Sandeep Kumar, skumar@engr.ucr.edu

**Author contributions**

The manuscript was written through contributions of all authors. All authors have given approval to the final version of the manuscript. ‡PCL, AK and RGB have equal contribution to this work.

**Conflict of interest**

Authors declare no conflict of interest.

**Acknowledgement**

We thank Nicola Spaldin for valuable discussions.

The fabrication of the experimental devices was done at the Center for Nanoscale Science and Engineering at UC Riverside. The electron microscopy sample preparation and imaging were done at the Central Facility for Advanced Microscopy and Microanalysis at UC Riverside.

SK acknowledges a research gift from Dr. S Kumar. D.M.J. is supported by the Swiss National Science Foundation (SNSF) under Project 184259.

# Supplementary Materials – Possible evidence of dynamical multiferroicity in flexoelectronic silicon at room temperature

**Materials and methods**

The devices were made using micro/nano fabrication methods. The sample and electrodes were patterned using photolithography. The device layer Si (wafer resistivity- 0.001-0.005 $\Omega$-cm) was then etched using deep reactive ion etching (DRIE). The samples were then made freestanding using the hydrofluoric acid (HF) vapor etch, as shown in Supplementary Figure S1. The HF vapor etched the $SiO_2$ underneath the patterned sample. Using an optical microscope, the HF vapor etch was stopped once the undercut in the oxide was sufficient to make the sample freestanding. Then, the MgO layer was deposited using an RF magnetron sputtering system. The metal layers (Pt or Py) were deposited using e-beam evaporation. E-beam evaporation was used since it gave a line-of-sight deposition as well as large residual stresses. In the case of the Py layer, we also deposited 1 nm of Pd to protect the Py layer from oxidation.

The ADMR measurement in the Pt/p-silicon sample was carried out using delta mode with a Keithley 6221 current source and a 2182A nanovoltmeter. The delta mode utilized the current reversal technique to cancel out any constant thermoelectric offsets, which was essential to minimize the thermal transport effects. The magnitude of the current used for measurement is 100 µA.

All the other measurements were carried out using an AC lock-in technique in the low frequency regime (5-37 Hz). The low frequency measurement lead to a quasi-static response and a time-dependent response is not recorded.

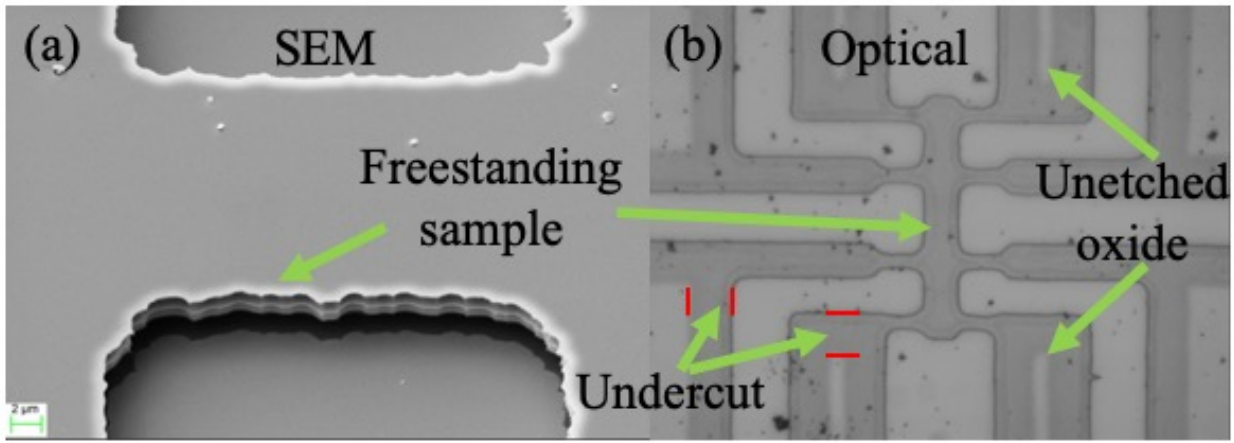

Supplementary Figure S1. (a) A scanning electron micrograph showing the freestanding nature of the sample (reflection). The oxide layer underneath the sample area is etched using an HF vapor etch. (b) An optical micrograph showing the etch contrast in the freestanding sample area and undercut at electrodes and connecting arms. The undercut in the oxide layer is essential for making the sample freestanding.

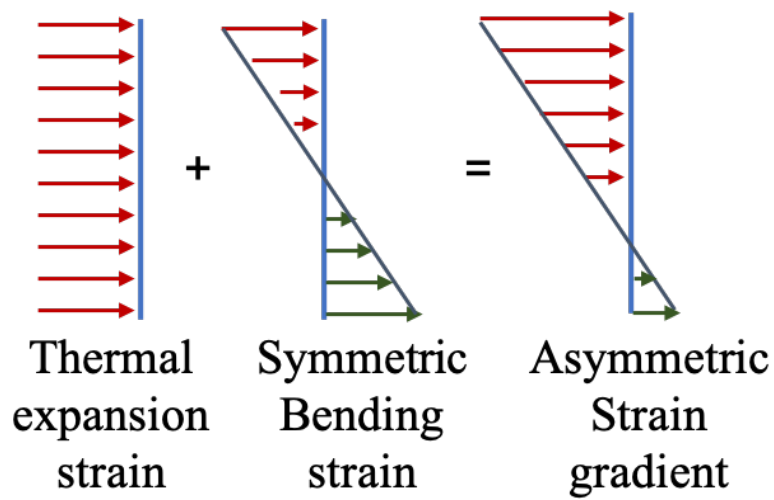

Supplementary Figure S2. A schematic showing the origin of the asymmetric strain gradient, which is the underlying cause of asymmetric electronic charge separation.

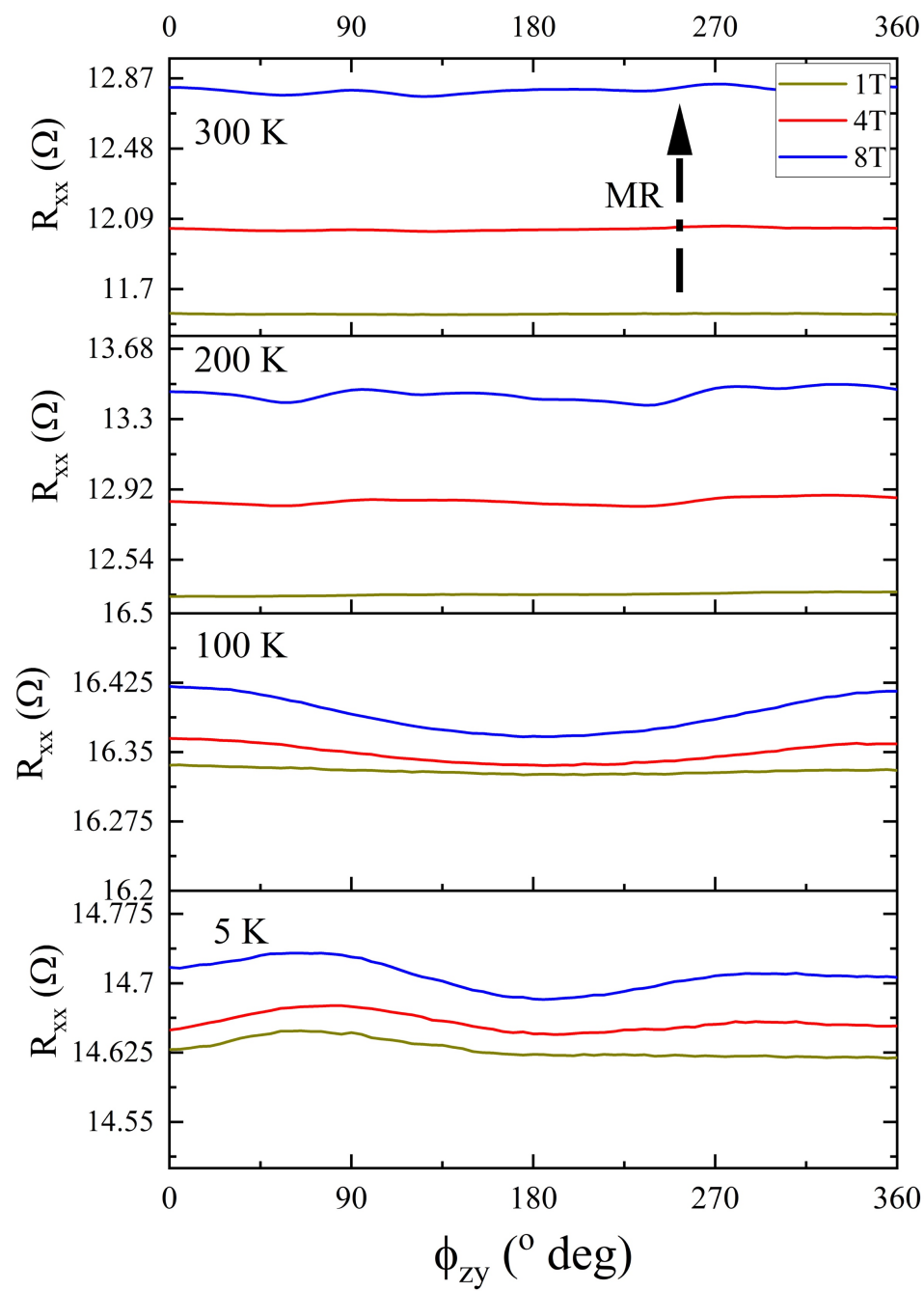

Supplementary Figure S3. The angle-dependent longitudinal resistance as a function of constant magnetic field of 1 T, 4 T and 8 T at 300 K, 200 K, 100 K and 5 K.

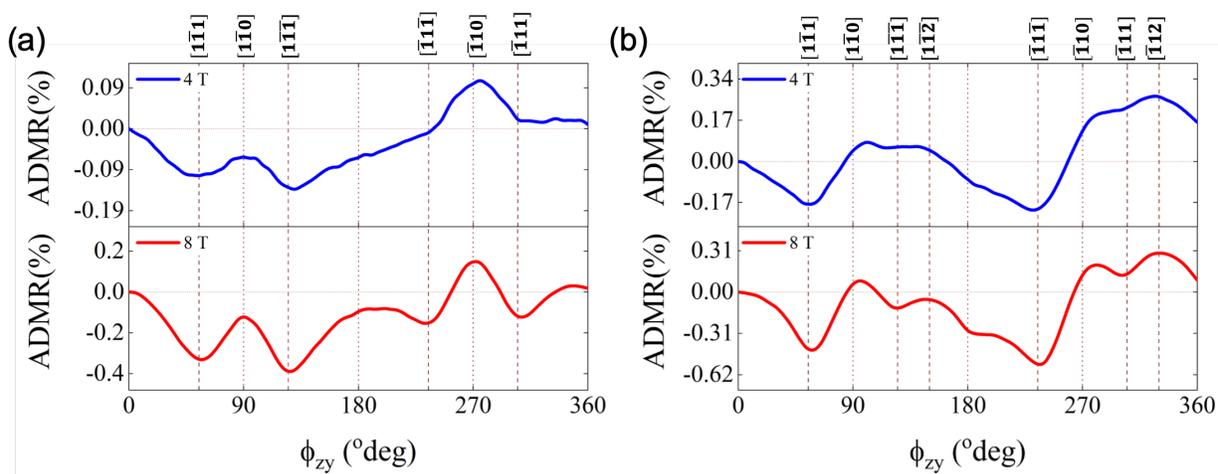

Supplementary Figure S4. The ADMR response at 300 K and 200 K showing the crystallographic direction-dependent anisotropy.

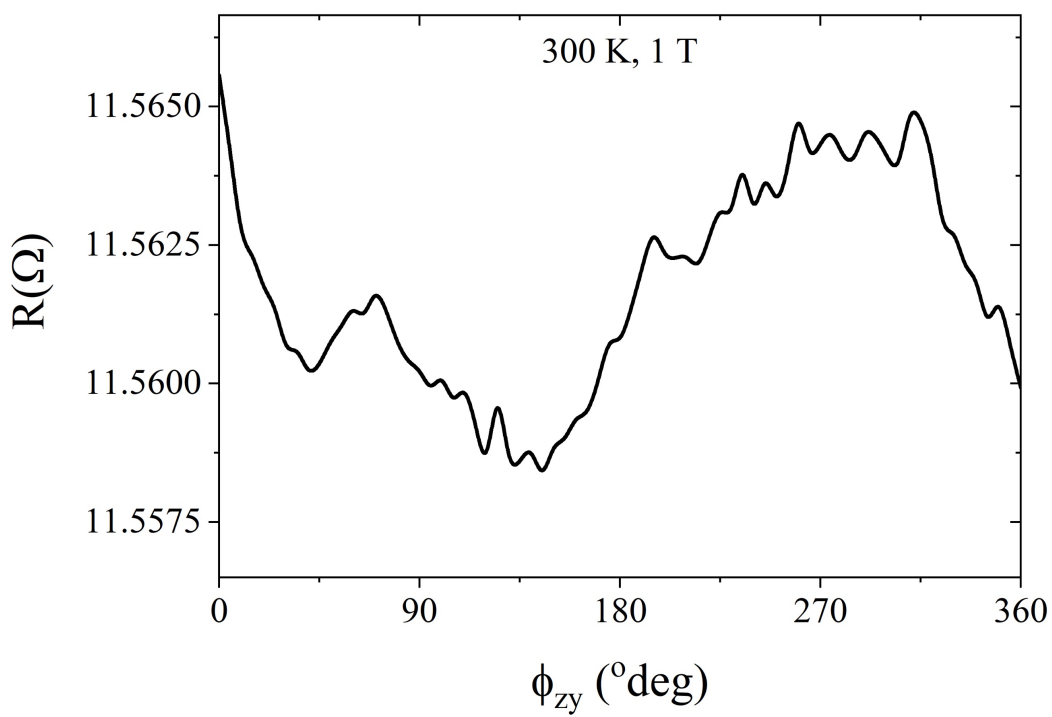

Supplementary Figure S5. The angle-dependent longitudinal resistance at 300 K and 1 T magnetic field showing weak direction dependent behavior.

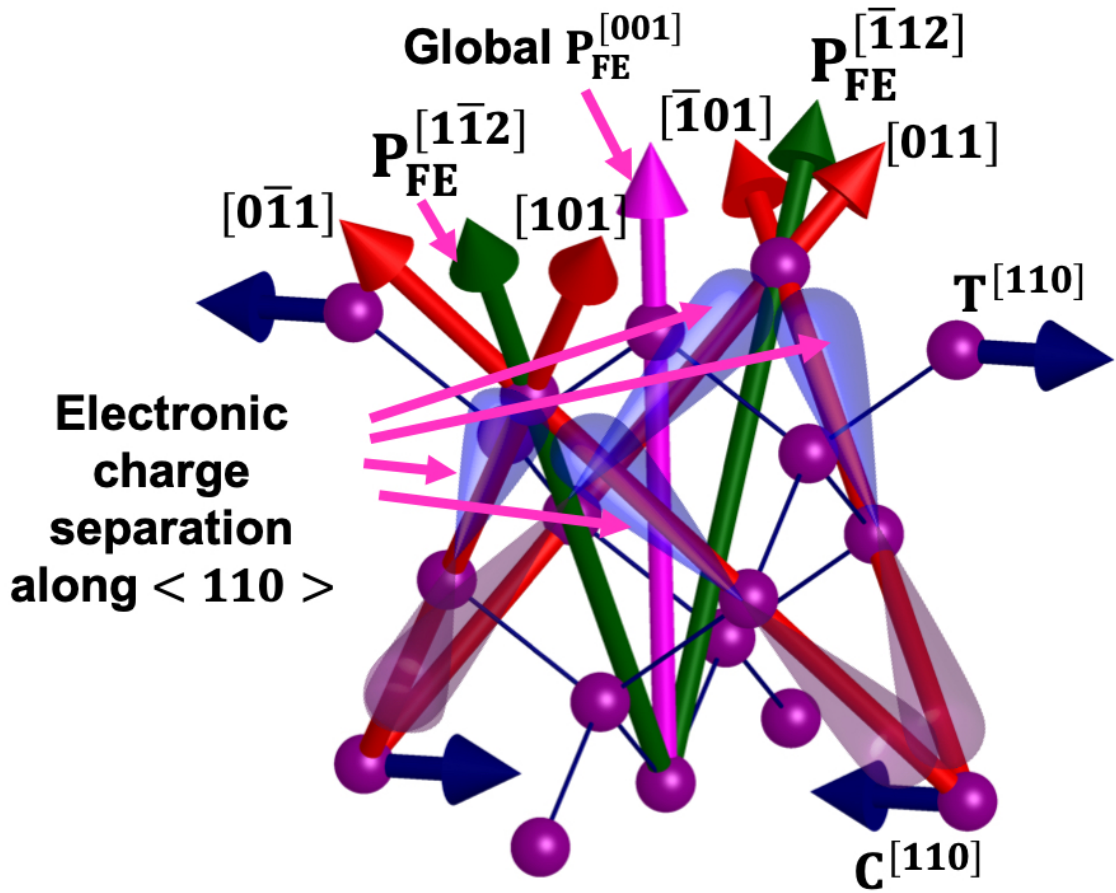

Supplementary Figure S6. A schematic showing the electronic charge separation along four $<110>$ directions and, as a consequence, local flexoelectronics polarization was along $[1\bar{1}\bar{2}]$ and $[\bar{1}12]$. The global flexoelectronics polarization was [001] and parallel to the strain gradient direction.

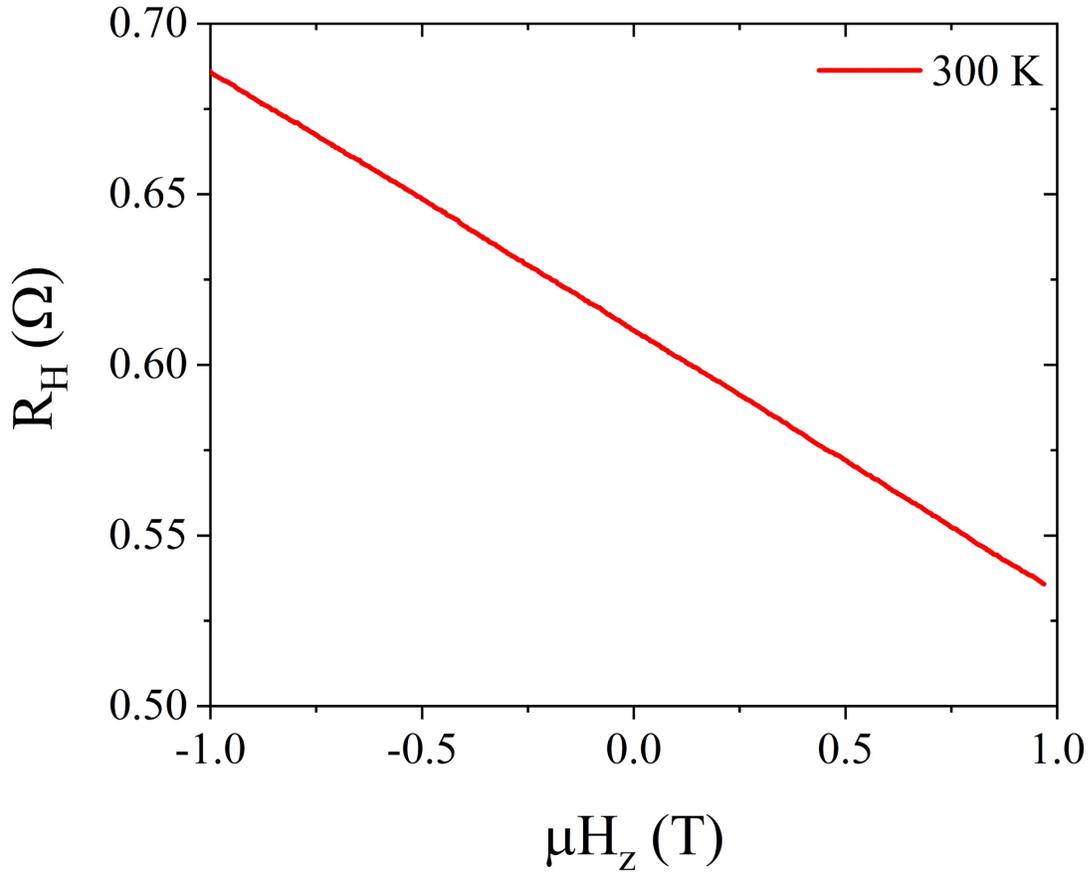

Supplementary Figure S7. The Hall effect measurement on an MgO (2 nm)/p-Si (2 μm) sample using 2 mA of current. In spite of p-doped Si, the Hall resistance showed a negative slope corresponding to the electrons being the charge carrier. This behavior might arise due to charge separation from the flexoelectronic effect.

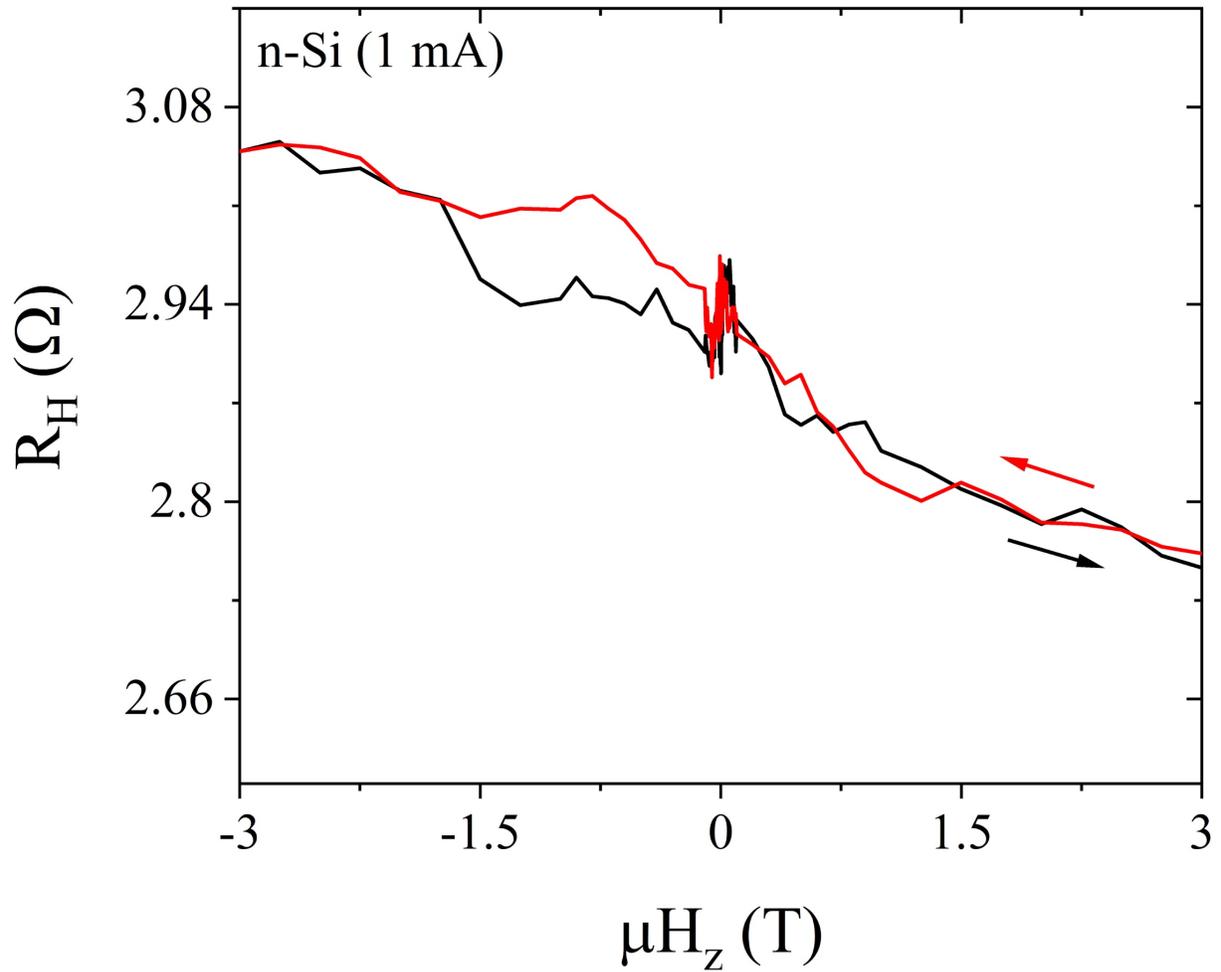

Supplementary Figure S8. The Hall effect measurement on an MgO (2 nm)/n-Si (2 μm) sample using 1 mA of current. As compared to measurement at 5 mA shown in main text, this measurement did not show AHE behavior explicitly. We attributed this behavior to a smaller strain gradient. The charge carrier concentration was estimated to be ~$5.9\times10^{19}$ cm$^{-3}$ at 1 mA current in this measurement.